\pdfoutput=1

\documentclass[11pt,twoside,a4paper,cmspaper,final,collab]{cms-tdr}
\def\svnVersion{361148}\def\cmsCernNoTag{CERN-EP/2016-041}\def\cmsCernDate{\today}\def\cmsMessage{Published in the European Physical Journal C as \href{http://dx.doi.org/10.1140/epjc/s10052-016-4206-6}{\doi{10.1140/epjc/s10052-016-4206-6.}}}

\begin{document}\cmsNoteHeader{EXO-12-053}

\hyphenation{had-ron-i-za-tion}
\hyphenation{cal-or-i-me-ter}
\hyphenation{de-vices}
\RCS$Revision: 354156 $
\RCS$HeadURL: svn+ssh://svn.cern.ch/reps/tdr2/papers/EXO-12-053/trunk/EXO-12-053.tex $
\RCS$Id: EXO-12-053.tex 354156 2016-07-09 06:36:58Z mgouzevi $
\newlength\cmsFigWidth
\ifthenelse{\boolean{cms@external}}{\setlength\cmsFigWidth{0.49\textwidth}}{\setlength\cmsFigWidth{0.8\textwidth}}
\ifthenelse{\boolean{cms@external}}{\providecommand{\cmsLeft}{top}}{\providecommand{\cmsLeft}{left}}
\ifthenelse{\boolean{cms@external}}{\providecommand{\cmsRight}{bottom}}{\providecommand{\cmsRight}{right}}
\newcommand{\intlumi}{19.7\fbinv}
\newcommand{\mjj}{\ensuremath{m_\mathrm{jj}}\xspace}
\newcommand{\mH}{\ensuremath{m_{\PH}}\xspace}
\newcommand{\mx}{\ensuremath{m_\mathrm{X}}\xspace}
\newcommand{\nsub}{\ensuremath{\tau_{21}}\xspace}
\newcommand{\etaj}{\ensuremath{\abs{\eta}}\xspace}
\newcommand{\deta}{\ensuremath{\abs{\Delta\eta_\mathrm{jj}}}\xspace}
\newcommand{\mjp}{\ensuremath{m_\mathrm{j}^\mathrm{P}}\xspace}
\newcommand{\LambdaR}{\ensuremath{\Lambda_\mathrm{R}}\xspace}
\newcommand{\HH}{\PH\PH}
\newcommand{\PX}{\ensuremath{\mathrm{X}}\xspace}
\cmsNoteHeader{EXO-12-053} % This is over-written in the CMS environment: useful as preprint no. for export versions
\title{Search for heavy resonances decaying to two Higgs bosons in final states containing four b quarks}

\date{\today}

\abstract{A search is presented for narrow heavy resonances X decaying into pairs of Higgs
  bosons (\PH) in proton-proton collisions collected by the CMS experiment
  at the LHC at $\sqrt{s}=8\TeV$. The data
  correspond to an integrated luminosity of 19.7\fbinv.
  The search considers $\PH\PH$ resonances with masses between 1 and 3\TeV, having final states of two b quark pairs.
  Each Higgs boson is produced with large momentum,
  and the hadronization products of the pair of b quarks can usually be reconstructed
  as single large jets.
  The background from multijet and \ttbar events is significantly reduced by applying
  requirements related to the flavor of the jet, its mass, and its substructure.
  The signal would be identified as a peak on top of the dijet invariant mass spectrum of the remaining background events.
  No evidence is observed for such a signal.
  Upper limits obtained at 95\% confidence level for the product of the production
  cross section and branching fraction $\sigma({\Pg\Pg} \to \PX)\,  \mathcal{B}({\PX} \to \PH\PH \to \bbbar\bbbar)$
  range from 10 to 1.5\unit{fb} for the mass of X from 1.15 to 2.0\TeV, significantly extending previous searches.
  For a warped extra dimension theory with a mass scale $\Lambda_\mathrm{R} = 1$\TeV, the data exclude radion scalar masses between 1.15 and 1.55\TeV.}
\hypersetup{%
pdfauthor={CMS Collaboration},%
pdftitle={Search for heavy resonances decaying to two Higgs bosons in final states containing four b quarks},%
pdfsubject={CMS},%
pdfkeywords={CMS, physics, dijet, jet substructure, resonances}}

\maketitle
\section{Introduction}
\label{sec:introduction}

The production of pairs of Higgs bosons ($\PH$) in the standard model (SM) has a
predicted cross section in gluon-gluon fusion at $\sqrt{s}=8\TeV$~\cite{deFlorian:2013jea,Baglio:2012np} for the Higgs boson mass $\mH \approx 125\GeV$~\cite{Aad:2015zhl} of only  $10.0 \pm 1.4\unit{fb}$. Many BSM theories suggest the existence of narrow heavy particles \PX~ that can decay to a pair of Higgs bosons~\cite{Djouadi:2005gj,Barbieri:2013nka, Randall:1999ee, Goldberger:1999uk,DeWolfe:1999cp,Csaki:1999mp,Davoudiasl:1999jd,Csaki:2000zn, Agashe:2007zd}.
The natural width for such a resonance is expected to be a few percent of its pole mass $\mx$, which corresponds to a typical detector resolution.
In contrast, the SM production of Higgs boson pairs results in a broad distribution of effective mass, falling mainly in the range from 300 to 600\GeV. Thus the presence of a narrow state would be readily detected, even if produced with a cross section as small as that for the SM process.

Searches for narrow particles decaying to two Higgs bosons have already been performed by the
ATLAS~\cite{Aad:2014yja, Aad:2015uka, Aad:2015xja} and  CMS~\cite{Khachatryan:2014jya, Khachatryan:2015year, Khachatryan:2015tha, Khachatryan:2016sey} collaborations
in \Pp\Pp~collisions at the CERN LHC. Until now their reach was limited to $\mx \leq 1.5\TeV$.
Because longitudinal \PW~ and \PZ~ states are provided by the Higgs field in the SM, any $\PH\PH$ resonance potentially also decays into $\PW\PW$
and $\PZ\PZ$ final states.
Searches for $\PX \to \PW\PW$, $\PZ\PZ$, and $\PW\PZ$ states were performed by ATLAS and CMS~\cite{ATLASVV, ATLASWV, ATLASZV, Khachatryan:2014hpa, CMSZVWV}.
The combinations of these results \cite{Aad:2015ipg, CMSZVWV, Brehmer:2015dan, Dias:2015mhm} indicate that the region
around $\mx \approx 2\TeV$ is particularly interesting to explore.

This paper reports on a search for $\PX \to \PH\PH$ covering the mass range  $1.15 < \mx < 3.0\TeV$,
significantly extending the reach of the present results beyond $1.5\TeV$.
The final state that provides the best sensitivity in this mass range is $\PH\PH \to \bbbar\bbbar$, which benefits from the expected large branching fraction ($\mathcal{B}$) of 57.7\% for $\PH \to \bbbar$~\cite{Heinemeyer:2013tqa} and a relatively low background from SM processes.

Many BSM proposals explicitly considered in this paper postulate the existence of a warped extra dimension (WED)~\cite{Randall:1999ee} and predict the
existence of a scalar radion~\cite{Goldberger:1999uk,DeWolfe:1999cp,Csaki:1999mp}. The radion is a spin-0 resonance associated with the fluctuations in the length of the extra dimension.
The production cross section as a function of $\mx$ is proportional to $1/\LambdaR^2$, where $\LambdaR$ is the scale parameter of the theory. In this paper we consider two cases: $\LambdaR = 1$ and $3\TeV$. In the first case, the WED theory predicts a cross section that can be detected at the LHC~\cite{Khachatryan:2015year}, but is challenged by the constraints derived from the electroweak precision measurements~\cite{Archer:2014jca}. This specific model is excluded up to $\mx = 1.1\TeV$ by the previous $\PX \to \PH\PH$ searches~\cite{Aad:2015uka, Khachatryan:2015year}. In contrast, the predicted cross section for $\LambdaR=3 \TeV$ is a factor of 9 times smaller, but the theory is less constrained by these searches.
We consider that the radion is produced exclusively via gluon-gluon fusion processes, with $\mathcal{B}(\text{radion} \to \HH) \approx 25\%$ above 1\TeV.

{\tolerance=1200
In the mass range of this search, the topology of the $\bbbar\bbbar$ final state is constrained by the size of the Lorentz boost of the Higgs bosons that is typically $\gamma_{\PH} \approx \mx / 2\mH \gg 1$ and defines the so-called boosted regime~\cite{Gouzevitch:2013qca,Cooper:2013kia, Butterworth:2008iy}. In this regime each Higgs boson is produced with a large momentum and its decay products are collimated along its direction of motion.
The hadronization of a pair of narrowly separated $\cPqb$ quarks will result in a single
reconstructed jet of mass compatible with $\mH$. The $\PH$ candidates are selected by employing jet substructure techniques to identify jets
containing constituents with kinematics consistent with the decay of a highly boosted Higgs boson.
These candidates are then required to be consistent with  decays of $\PB$~hadrons, based on our
\PQb tagging algorithms. The signal
is identified in the dijet mass ($\mjj$) spectrum as a peak above a falling background which originates mainly from multijet events and $\ttbar$ production.
\par}

\section{The CMS detector}
\label{sec:cms_detector}

The central feature of the CMS apparatus is
a superconducting solenoid of 6\unit{m} internal diameter, providing a
magnetic field of 3.8\unit{T}. A
silicon pixel and strip tracker, a lead tungstate crystal electromagnetic
calorimeter, and a brass and scintillator hadron calorimeter,
each composed of a barrel and two endcap sections, reside within the solenoid volume.
Extensive forward calorimetry complements
the coverage provided by the barrel and endcap detectors.
Muons are
measured in gas-ionization detectors embedded in the steel flux-return
yoke outside the solenoid. A detailed description of the CMS detector, together with a definition
of the coordinate system and the basic kinematic variables, can
be found in Ref.~\cite{Chatrchyan:2008zzk}.

\section{Simulated events} \label{sec:data_and_mc_samples}

Monte Carlo (MC) simulations are used to provide: predictions of background processes, optimization of the event selection, and cross-checks of data-based background estimations.

Signal, multijet and $\ttbar$ background events are generated using the leading-order
matrix element generator
\MADGRAPH 5v1.3.30~\cite{Madgraph_Alwall:2011uj, Madgraph_Alwall:2011uj}.
Parton shower and hadronization are included using \PYTHIA 6.4.26~\cite{Sjostrand:2006za},
and the matrix element is matched to the parton shower using the MLM scheme~\cite{Hoche:2006ph}.
The Z2* tune is used to describe the
underlying event. This tune is identical to the Z1 tune~\cite{tunesZ2}, but uses
the CTEQ6L parton distribution functions (PDF)~\cite{Pumplin:2002vw}. The signal events are simulated with an intrinsic width of the radion
fixed to 1\GeV, $\mH=125\GeV$.
Different samples are generated for $\mx$ ranging from 1.15 to 3 \TeV.
All generated events are processed through a simulation of the CMS
apparatus based on \GEANTfour~\cite{Geant4}.
Additional $\Pp\Pp$ interactions within a bunch crossing (pileup) are added 
to the simulation, with a frequency distribution chosen to match that observed in data. During this data-taking period the 
mean number of interactions per bunch crossing is 21.

\section{Event reconstruction and selections}\label{sec:analysis}

The analysis is based on data from $\Pp\Pp$ interactions observed with the CMS
detector at $\sqrt{s}=8\TeV$. The
data correspond to an integrated luminosity of
\intlumi.
Events are collected using at least one of the two specific trigger conditions based on jets reconstructed
online: the first trigger requires a large $\mjj$ calculated for the two jets of highest transverse momentum (referred to as leading jets);
the second trigger requires a
large value of $\HT = \sum_i \pt^i$, where the sum runs over
the reconstructed jets in the event with transverse momenta $\pt>40\GeV$. The lower
thresholds applied to $\mjj$ and the $\HT$ triggers were changed during the
data-taking period to maintain a constant trigger rate while the LHC peak luminosity steadily
increased. More than half of the data were collected with
$\mjj > 750\GeV$ and $\HT > 650\GeV$. The remaining data were collected with the requirement $\HT>750\GeV$.

Events are required to have at least one reconstructed $\Pp\Pp$ collision vertex within $\abs{z} < 24\unit{cm}$ of the center of the detector along the longitudinal
beam directions.
Many additional vertices, corresponding to pileup interactions, are usually reconstructed in an event using charged particle tracks.
We assume that the primary interaction vertex corresponds to the one that maximizes the sum in $\pt^2$ of
these associated tracks.

Individual particles are reconstructed using a particle-flow (PF)
algorithm~\cite{PFPAS2009, CMS-PAS-PFT-10-001} that combines the information from all the CMS detector
components. Each such reconstructed particle is referred to as a PF candidate.
The five classes of PF candidates correspond to muons, electrons,
photons, and charged and neutral hadrons. Charged hadron candidates
not originating from the primary vertex of the event are discarded to reduce contamination from pileup~\cite{CMS-PAS-JME-14-001}.

The Cambridge--Aachen (CA) algorithm~\cite{CAaachen},  implemented in \textsc{FastJet}~\cite{Cacciari:2011ma},
clusters PF candidates into jets using a distance parameter $R = 0.8$.
An event-by-event jet area-based
correction~\cite{jetarea_fastjet,jetarea_fastjet_pu,CMS-PAS-JME-14-001} is applied
to each reconstructed jet to remove the remaining energy originating from pileup vertices
primarily consisting of neutral particles.  The jet four-momenta are also
corrected to account for the difference between the measured and the expected momentum at the particle level,
using the standard CMS correction procedure described in Refs.~\cite{JINST6,CMS-DP-2013-011}.

Events are required to have at least two jets, and the two leading jets each to have $\pt> 40\GeV$
and pseudorapidity $\etaj < 2.5$. In addition, identification criteria
are applied to remove spurious jets associated with calorimeter noise~\cite{PFPAS2009}. To reduce the contribution
from multijet events, the two leading jets must be relatively close in $\eta$, $\deta<1.3$,
a selection discussed in Refs.~\cite{CMS:2012yf, Khachatryan:2014hpa}. Events with $\mjj<1\TeV$ are rejected. Above this mass threshold,
the efficiency of the trigger requirement for the chosen selections exceeds 99.5\%.

\begin{figure}[th!b]
\centering
\includegraphics[width=0.49\textwidth]{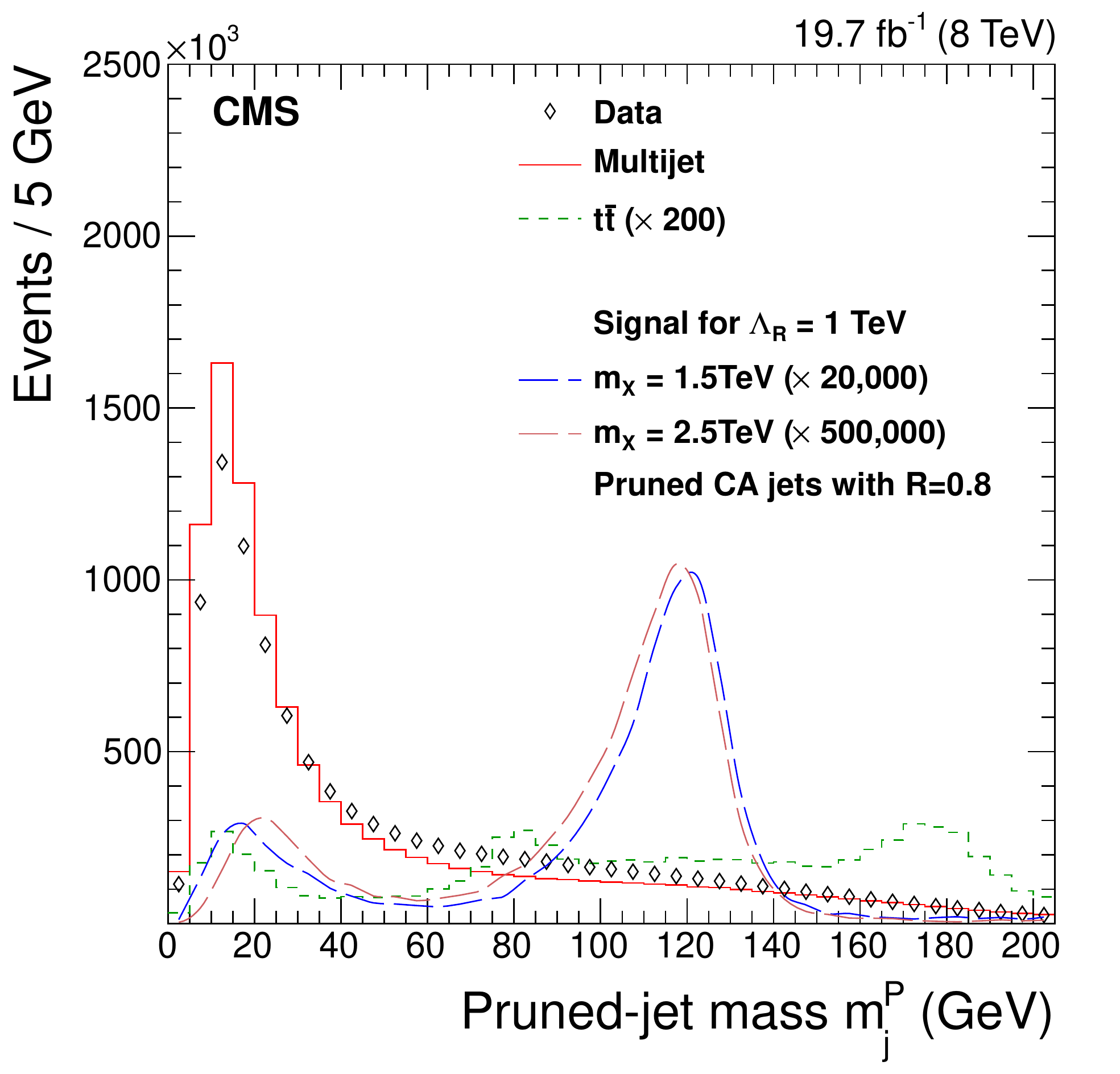}

\caption{Simulated $\mjp$ spectrum for spin-0 radion signals, multijet and \ttbar events, and the
spectrum measured in data. Each event contributes twice to the distribution, once per jet. The multijet
contribution is rescaled to match the event yield in data, while the signal and \ttbar spectra are rescaled
by the large factors indicated, to be visible in the figure.\label{fig:taggingvariables}}
\end{figure}

The mass and \PQb flavour properties of the leading jets are used to suppress the multijet and \ttbar backgrounds.
Soft gluon radiation and a fraction of the remaining neutral pileup particles are first removed from each jet through the implementation of a jet-grooming algorithm called jet pruning~\cite{jetpruning1,jetpruning2}.
This technique reduces significantly the mass of jets originating from quarks and gluons \cite{Chatrchyan:2013vbb}, while improving the resolution of the jets resulting from the hadronic decays of a heavy SM boson \cite{Khachatryan:2014vla}.
The invariant mass $\mjp$ is calculated for the two leading pruned jets.
In Fig.~\ref{fig:taggingvariables}, the $\mjp$ distribution of the two leading jets is shown
for data, signal, and background events.
For jets initiated by a quark or a gluon, $\mjp$ peaks around 15\GeV, while jets from
high-momentum Higgs boson decay usually have a pruned mass around 120\GeV.  The difference of $\approx 5\GeV$ relative to the nominal $m_\PH$ value is related to the presence of neutrinos produced by the semileptonic
decays of \PB~mesons, and the inherent nature of  the pruning procedure.
A small peak near 15\GeV is also observed for signal events, and
corresponds mainly to asymmetric decays in which the jet pruning algorithm removes the decay products of one of the two \PB~mesons.
Each of the leading jets has to satisfy $110 <\mjp<135\GeV$, a requirement that is chosen to maximize
the sensitivity of the analysis to the presence of a narrow resonances.
Some differences are observed between the data and background estimated from simulation. 
These discrepancies do not affect the results of this analysis since the background is estimated using techniques based on data only.

The identification of jets likely to have originated from the hadronization of a pair of \PQb quarks exploits
the combined secondary vertex (CSV) \PQb jet tagger~\cite{Chatrchyan:2012jua}. This algorithm combines the
information from track impact parameters and secondary vertices within a given jet into a continuous output discriminant
~\cite{Chatrchyan:2012jua, CMS-PAS-BTV-13-001}.
The working point used in this paper corresponds to an efficiency of 80\% for identifying b jets and a rate of 10\% for mistagging jets from light
quarks or gluons as originating from \PQb quarks. This working point was chosen to maximize the sensitivity of the analysis, while retaining a sufficient number of events to allow a reliable estimation of the background.

In the first step of the procedure used to select \PH jet candidates, the pruned jets are split into two subjets by reversing the final iteration in the jet clustering algorithm.
The angular separation between the subjets is $\Delta R \equiv \sqrt{\smash[b]{(\Delta \eta)^2 +(\Delta \phi)^2}}$, where $\eta$ is the pseudorapidity and $\phi$ the azimuthal angle. Two cases are considered, with the transition between them occurring at $\mx \approx 1.6\TeV$:

\begin{enumerate}
\item $\Delta R > 0.3$: in this group the jet is considered to be \PQb tagged
if at least one subjet satisfies the requirements of the CSV
working point. Moreover, the jet is considered as ``double \PQb tagged" if
both subjets satisfy the CSV requirement.

\item $\Delta R<0.3$: here the subjet \PQb tagging selection is inefficient~\cite{CMS-PAS-BTV-13-001}.
The \PQb tagging algorithm is therefore applied directly to the jet. In this case it is not possible to distinguish
between \PQb-tagged and double \PQb-tagged jets, and therefore either of these two possibilities are accepted.
\end{enumerate}

In summary, a jet is considered an $\PH$ jet candidate
if it satisfies the mass and \PQb tagging requirements.
Events are selected when both leading jets are $\PH$ jets,
and at least one of them is double \PQb tagged.
The simulated results are corrected to match the $\PH$ and \PQb tagging efficiencies observed in data~\cite{CMS-PAS-BTV-13-001}.

A final selection is based on the kinematic properties of the constituents of $\PH$ jets.
The quantity N-subjettiness~\cite{Thaler:2010tr,
  Thaler:2011gf, Stewart:2010tn} $\tau_N$ is used to quantify the degree to which
constituents of a jet can be arranged into $N$ subjets.
The ratio $\nsub = \tau_2/\tau_1$  is calculated for each of the two $\PH$ jet candidates.
High- (HP) and low-purity (LP) Higgs boson
candidates are defined as having $\nsub < 0.5$ and $0.5 \leq \nsub < 0.75$,
respectively. Events are required to have at least one HP $\PH$ jet
and another $\PH$ jet that passes either the HP or LP requirements.

The sample of events satisfying the previously defined criteria is subsequently divided into
three categories. Events with two high-purity $\PH$ jets form the
HPHP category. Among the remaining events, those for which the
high-purity $\PH$ jet is the leading jet constitute the
HPLP category. The rest of the sample constitutes the LPHP
category.

The selection criteria applied to reduce the background are summarized in Table~\ref{table:selections}.
The region of phase space defined by all these criteria is referred to as the signal region.
The fraction of the simulated signal and \ttbar samples, satisfying these criteria, as well as the number of data events
passing the selections is also provided.

\begin{figure}[th!b]
\centering
\includegraphics[width=0.49\textwidth]{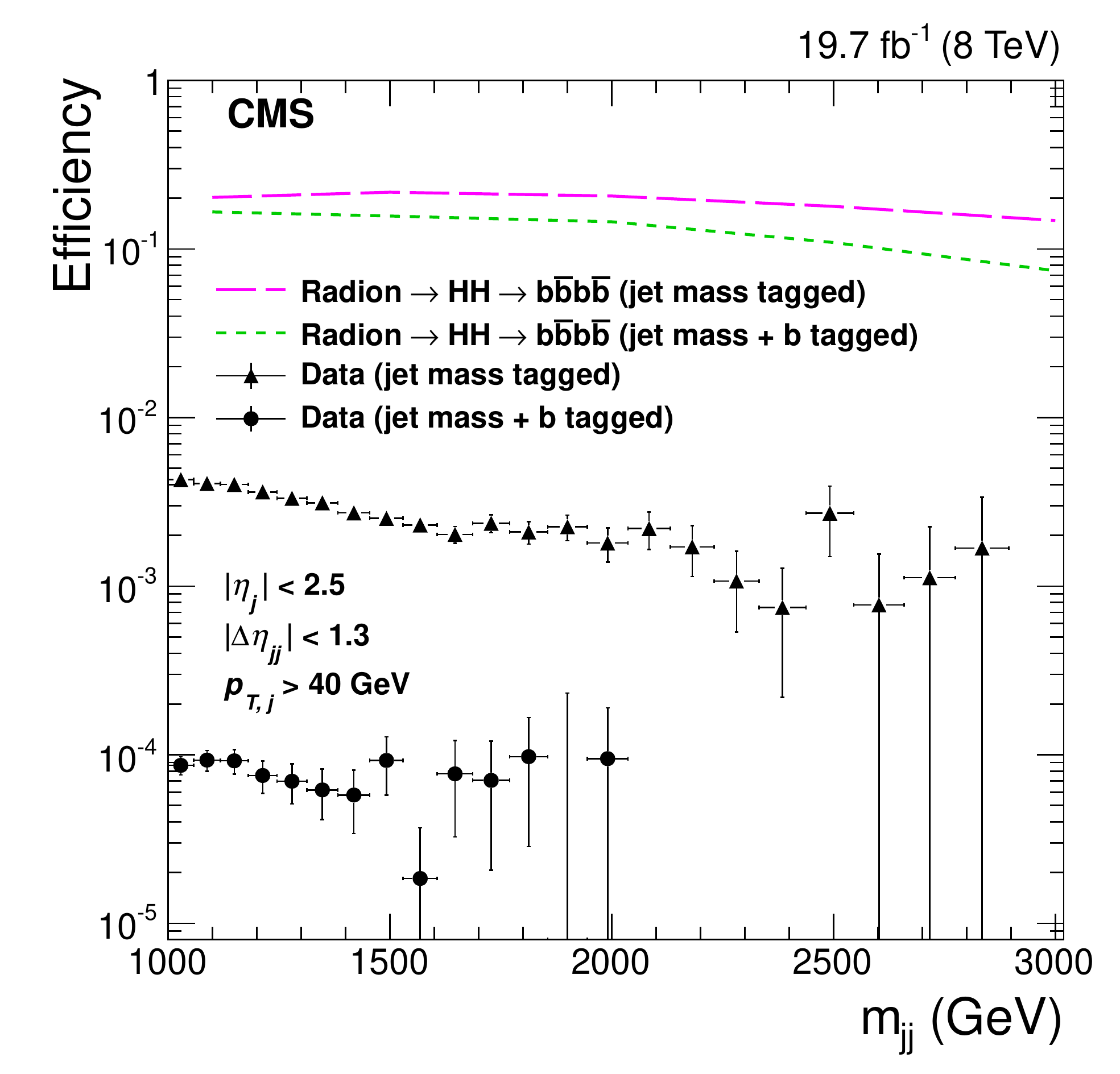}
\caption{
The efficiencies of $\PH$ mass requirement and combined $\PH$ mass and \PQb tagging criteria, for data and signal.
  Events are required to be in
  the fiducial region ($\etaj < 2.5$, $\pt >40\GeV$ for both jets and
  $\deta<1.3$). The horizontal bar on each data point indicates the width of the bin. \label{fig:doubleefficiencies}}
\end{figure}

The fiducial selection is defined by the two leading jets having $\etaj < 2.5$, $\pt >40\GeV$, and a separation $\deta<1.3$.
The fraction of the signal within this fiducial region depends on its
spin, and is $\approx$60\% for a spin-0
resonance.
The efficiency of the combined \PH mass and \PQb tagging criteria for 
events within the fiducial region, for signal and data, is shown in Fig.~\ref{fig:doubleefficiencies}. 
The number of data events is reduced by four orders of magnitude while the signal efficiencies range from 10 to 20\% with a weak dependence on $\mx$, and are observed to be independent of the spin of the resonance. Finally, the total acceptance times efficiency is
provided in Table~\ref{table:selections}, and varies between 4.0 and
8.8\%, with the largest fraction of events populating the HPHP category.

Figure~\ref{fig:doubleefficiencies} shows that the probability of
incorrectly identifying multijet or $\ttbar$ events as events with two
Higgs bosons is less than 0.1\%, and appears to be independent of
$\mjj$ within statistical uncertainties. A more precise quantification
is provided in Table~\ref{table:selections} for \ttbar events. In
particular, we observe that the dijet mass, the pruned jet mass, and \PQb
tagging criteria are each sufficient for reducing the \ttbar
background by an order of magnitude. In contrast, the
N-subjettiness criterion is inefficient in reducing it.

\begin{table*}[thb]
\centering
\topcaption{Summary of selection requirements, with their signal
  and \ttbar background efficiencies, and total number of
  events observed in data. The selection criteria are applied
  sequentially and the efficiencies are cumulative, except in the last
  section of the table dedicated to categorization.\label{table:selections}}
\def\arraystretch{1.3}
\begin{tabular}{lrrrlr}
  \hline
   \multicolumn{1}{c}{\multirow{3}{*}{Selection criteria}} & \multicolumn{4}{c}{Efficiency for} & \multicolumn{1}{c}{Observed} \\ \cline{2-5}
                                   &  \multicolumn{3}{c}{signal with $\mx$ (\TeVns)} & \ttbar & \multicolumn{1}{c}{events}\\\cline{2-4}
                             & 1.3 & 2.0 & 3.0 & & \\
  \hline
  \multicolumn{1}{c}{Fiducial acceptance} & & & & & \\ \cline{1-1}

  At least 2 jets with $\pt>40\GeV$, & & & & & \\
  $\etaj<2.5$, and $\deta < 1.3$ & 63\% & 61\% & 59\% & 29\% &\\
  \hline
   \multicolumn{1}{c}{Analysis selections} & & & & & \\ \cline{1-1}

  $\mjj > 1\TeV$ & 59\% & 59\% & 58\% &  3.5\% & 2\,677\,308 \\
  2 jets with  $110 <\mjp<135\GeV$ & 12\% & 12\% & 8.5\% & 0.29\% & 9\,977 \\
  2 \PQb-tagged jets and & & & & & \\
  $\geq$1 double \PQb tagged jets & 9.0\% & 8.5\% & 4.5\%  & 0.05\% & 217 \\
  2 jets with $\nsub < 0.75$ and & & & & & \\
  $\geq$1 jet with  $\nsub < 0.5$ & 8.6\% & 8.1\% & 4.0\% & 0.04\% & 162 \\
  \hline
   \multicolumn{1}{c}{Categorization} & & & & & \\ \cline{1-1}

  HPHP           & 6.3\% & 5.5\% & 2.4\% & 0.03\% & 63 \\
  HPLP           & 1.1\% & 1.2\% & 0.9\% & 0.007\% & 48 \\
  LPHP           & 1.2\% & 1.4\% & 0.7\% & 0.004\% & 51 \\
  \hline
\end{tabular}
\end{table*}

\section{Signal extraction}
\label{sec:signal_extraction}

The signal is identified in the binned $\mjj$ spectrum in bin widths chosen to match the resolution
of the dijet mass, as described in
Ref.~\cite{Chatrchyan:2011ns}. This resolution is $\approx$50\GeV
at $\mx = 1.15\TeV$, increasing slowly to $\approx$100\GeV for $\mx = 3\TeV$.

The analysis defines a likelihood, for each $\mx$ hypothesis, based on
the total number of events in data, signal, and background counted in
a mass window in each category.  These mass windows have a typical size of
three or four bins centered approximatively around $\mx$ (see Table~\ref{tab:CountingMassWindow})
and contains more than 95\% of signal events.
The amount of signal is
estimated in the mass window using MC simulation, while the amount of
background is estimated as the integral of a parameterized model.
The total likelihood combines the information from the
three event categories.

\begin{table}[th!b]
\topcaption{Mass windows used for different signal hypotheses.\label{tab:CountingMassWindow}}
\centering
\begin{tabular}{  c  c | c  c }
\hline
$\mx$  & Mass window  & $\mx$  & Mass window  \\
 (\GeVns{})&(\GeVns{})&(\GeVns{})&(\GeVns{})\\\hline
1150 & [1058, 1246] & 1700 & [1607, 1856] \\
1200 & [1118, 1313] & 1800 & [1687, 1945]\\
1300 & [1181, 1455] & 1900 & [1700, 2037]\\
1400 & [1313, 1530] & 2000 & [1856, 2132]\\
1500 & [1383, 1607] & 2500 & [2231, 2775]\\
1600 & [1455, 1770] & 3000 & [2775, 3279]\\
\hline
\end{tabular}
\end{table}

\section{Parameterization of background}
\label{sec:background}

After event selection, $\approx$75,~90, and 95\% of the total
background is expected to originate from multijet events in HPHP, HPLP, and LPHP
categories, respectively.  The remaining contribution is from \ttbar
production, which is modelled in simulation, and rescaled to the total
next-to-next-to-leading order cross section~\cite{Czakon:2013goa}. All
other backgrounds containing Higgs bosons or \PW/\PZ bosons decaying
into jets represent less than 1\% of the total background.

The total background is estimated from data, without separating the
multijet or \ttbar fractions.  The expected $\mjj$ background spectrum
is approximated by a falling exponential for $1 < \mjj < 3\TeV$,
\begin{equation}
\frac{\rd N_\text{Background}}{\rd \mjj} = N_B\, a\, \re^{-a(\mjj-1000\GeV)},
\label{eqBkg}
\end{equation}
where the parameterization has been chosen to minimize the correlation between the
normalization $N_B$ and slope $a$. We obtain $a$ from a fit to the
$\mjj$ distribution in a control region, defined as the portion of
phase space where one of the jets satisfies $110 < \mjp < 135\GeV$ and
the other jet is required to have $60 < \mjp < 100\GeV$.  This choice
of the window for $\mjp$ results from a compromise between limited signal
contamination, sufficiently large statistics, and similarity in
substructure properties between the sideband jet and the \PH jet.
To use this control region we assume
that there is no resonant signal in the Z\PH final state.

\begin{figure*}[hbt]
\centering
\includegraphics[width=0.49\textwidth,angle=0]{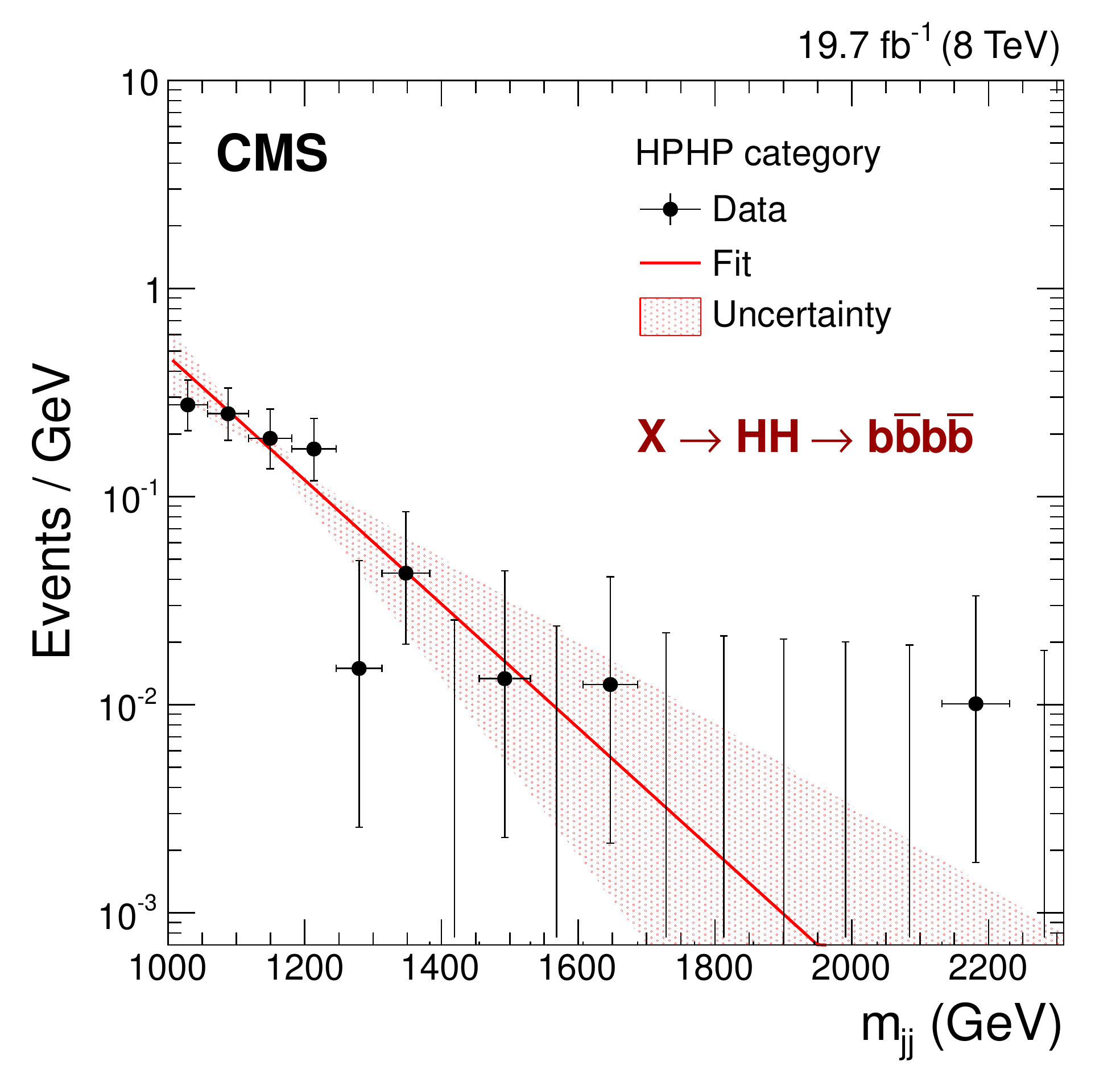}\\
\includegraphics[width=0.49\textwidth,angle=0]{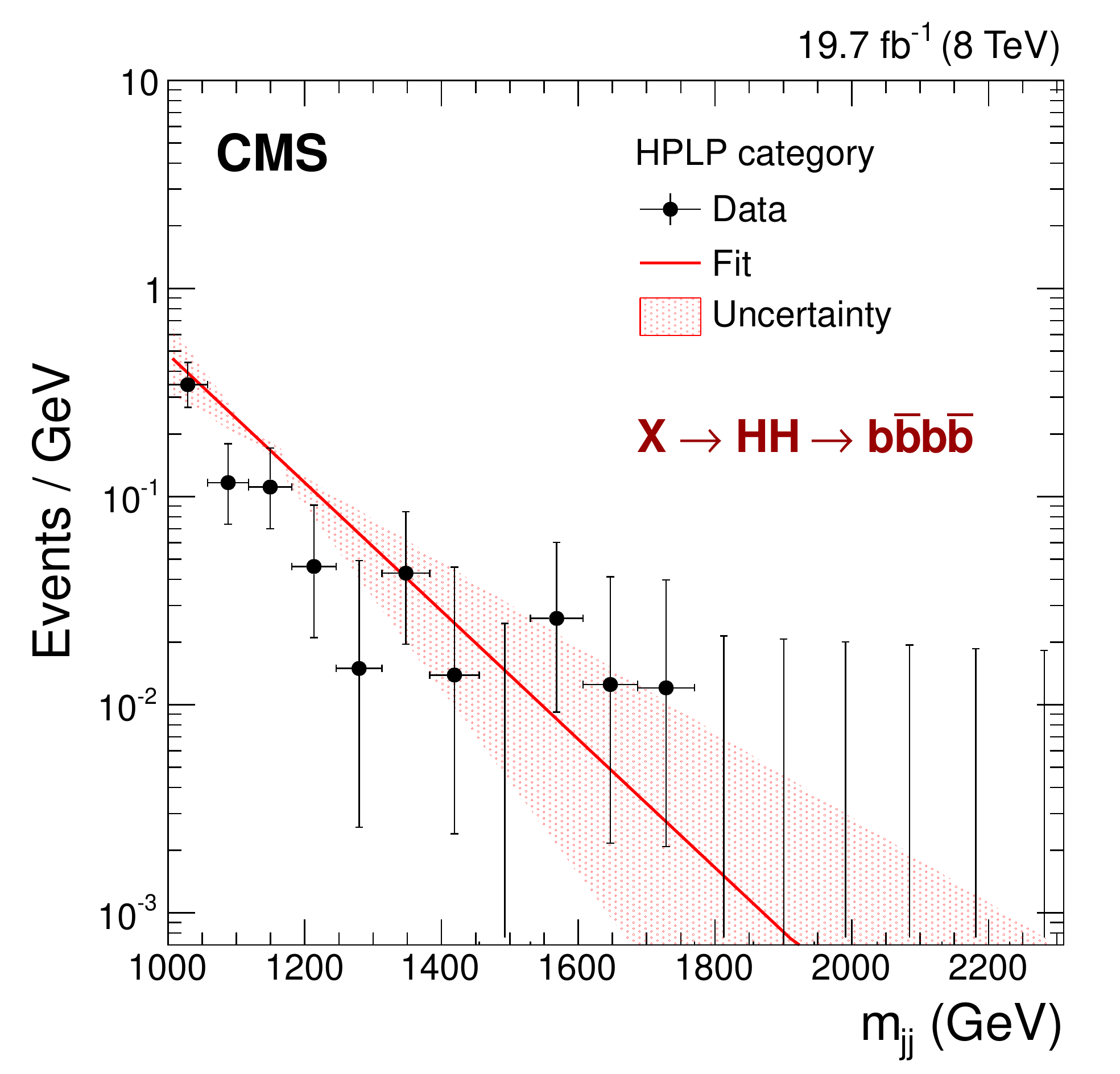}
\includegraphics[width=0.49\textwidth,angle=0]{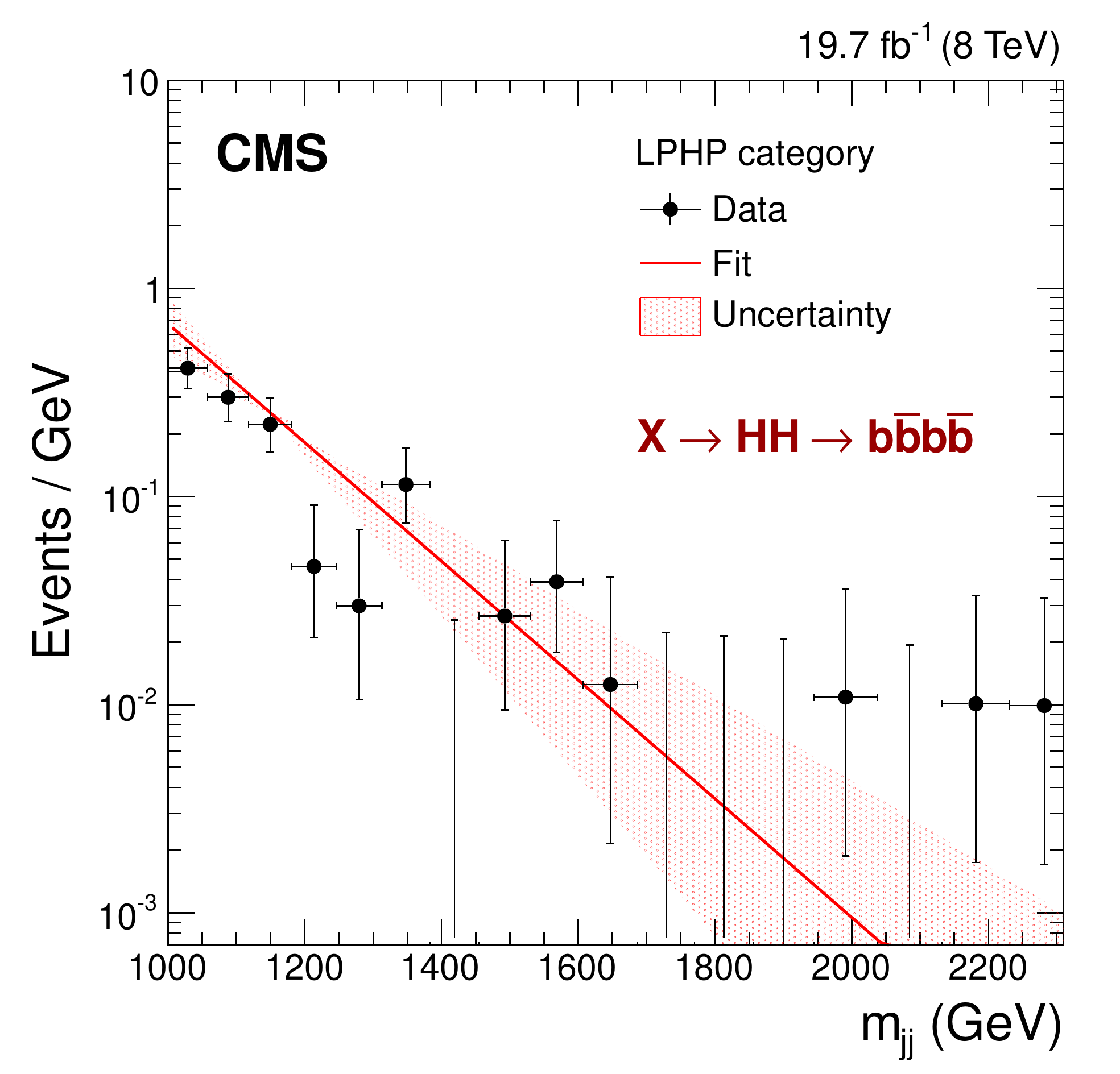}
\caption{Observed $\mjj$ spectrum (black points) in the control region
  together with the superimposed background fit (red line) and the
  uncertainty associated with the variation of the slope parameter $a$
  (red shaded area) for HPHP (top), HPLP (bottom-left), and LPHP
  (bottom-right) categories.  }
\label{fig:doubleVtagBG_RegionB}
\end{figure*}

The
control region contains between 1.1--2 times the number of events in the
signal region depending on the category. The result of the fit and the
uncertainty band associated with the uncertainty in the parameter $a$
are shown in Fig. \ref{fig:doubleVtagBG_RegionB}.
The effect of a residual contamination of the control region by the signal is
explicitly checked by adding an $\HH$ signal to the control region at different masses,
with a typical $\sigma(\Pg\Pg \to \PX \to \HH) \, \mathcal{B}(\PX \to \HH \to \bbbar\bbbar)$, corresponding to the sensitivity of the analysis at a given $\mx$.
The change in the slope parameter $a$ is observed to be negligible.

We extract $N_B$ for each signal hypothesis from the fit to the data that excludes events in the counting
window described in Section  \ref{sec:signal_extraction}. This background extraction procedure motivates the choice of
the lower value of the $\mx$ window for which the search is performed.
In order to improve the constraint on $N_B$, there must be at least one bin on the left side of the mass window to be retained.

This background estimation procedure assumes, on the one hand, that the $\mjj$ spectrum is
similar in the signal and the control regions, and on the other hand,
that it is similar for multijet and \ttbar event samples.  The
following cross-checks are performed to validate these hypotheses:

\begin{itemize}
\item The similarity of distributions for the signal and control
  regions are confirmed in the simulated multijet sample.
\item The parameters $a$ and $N_B$ are extracted from the signal
  region (using an approach similar to that of
  Ref.~\cite{Khachatryan:2014hpa}), and found to be compatible within
  statistical uncertainties with the parameters obtained through the
  normal method of background estimation.
\item The bin-by-bin normalization between the signal and control regions
is calculated using a sideband obtained by inverting the \PQb tagging criterion
on one of the jets (using a technique similar to that in Ref.~\cite{Chatrchyan:2012rva}),
and the normalization factor found to be independent of $\mjj$, within the statistical uncertainties.
\item The $\ttbar$ contribution in the signal region obtained from
  simulation is fitted by the function in Eq. (\ref{eqBkg})
and the resulting fit is found to be consistent with the distribution of the overall background within the statistical uncertainties.
\end{itemize}

Closure checks of the background-estimation procedure are performed
using simulated multijet events. These are also performed directly in
data in the control region. For this purpose, the control region is
split in two, a low mass control region with $60<\mjp<90\GeV$, and a
pseudo-signal region with $90<\mjp<100\GeV$. In both cases, the
predicted background is found to be compatible with that observed,
within the statistical uncertainties.

\section{Systematic uncertainties}
\label{sec:systematics}

The largest contributions to the systematic uncertainty in the signal
yields are the uncertainties associated with the classification of the
events into the purity categories, the estimation of the efficiency to
identify a \PH jet, and the calculation of the total integrated luminosity
(2.6\%)~\cite{CMS-PAS-LUM-13-001}, as well as with the determination of the jet energy scale
(JES) and resolution (JER). The major systematic uncertainties are
summarized in Table \ref{tab:Systematics}.

\begin{table*}[th!b]
\topcaption{
Typical uncertainties in different categories.\label{tab:Systematics}}
\centering
\begin{tabular}{  l| l {c}@{\hspace*{1pt}} c }
\hline
 \multicolumn{2}{c}{Source}  && Uncertainty \\
\cline{1-2}\cline{4-4}
 \multicolumn{2}{c}{\rule{0pt}{4ex} Background (statistical)} && 15 -- 100\%\\[1ex]
  \multirow{5}{*}{Signal (systematic)} & Luminosity && 2.6\%\\
 & \PQb tagging  && 3.8--14.4\%\\
 & Mass tagging && $5.2 \oplus 3.0\%$\\
 & JES $\oplus$ JER && $1.0 \oplus 1.0\%$ \\
 & Categorization &&  $^{+25}_{-19}\%$ (HPHP), $^{+59}_{-37}\%$ (HPLP), $^{+59}_{-37}\%$ (LPHP) \\[0.4ex]
\hline
\end{tabular}
\end{table*}

The uncertainty in the \PQb tagging efficiency originates from the uncertainty in the
data-to-simulation scale factors that are applied to the simulated
signal~\cite{CMS-PAS-BTV-13-001}. The scale factors are $\approx$90\% with an absolute
uncertainty between $\pm$3.8\% and $\pm$14\%, depending on the value of
$\mx$. The uncertainty increases at large $\mx$ because of the limited
amount of data available to constrain the scale factors.

The uncertainty in the mass selection efficiency is 2.6\% for each jet
and 5.2\% for the event.  This uncertainty is estimated by studying high \pt $\PW$
bosons in a $\ttbar$ data control sample~\cite{Khachatryan:2014vla}
and comparing to MC predictions. It includes the effect of the
difference in fragmentation between light and $\PQb$ quarks.
This uncertainty is fully correlated for all $\PH$ jets.  In
addition, the impact of the pileup modelling uncertainty in the Higgs boson
mass-tagging efficiency is assumed to be 1.5\% per jet,
\ie, 3\% for the event~\cite{Khachatryan:2014hpa}.

An uncertainty accounting for possible migration of signal events from
the HPHP to the HPLP and LPHP categories
results in uncertainties of $+25\%$ and $-19\%$, and of $+59\%$ and $-37\%$
in the normalization of the HPHP category, and of both the HPLP and LPHP categories, respectively.
These uncertainties are estimated by comparing
the $\nsub$ distribution in measured and simulated \ttbar events
~\cite{Khachatryan:2014vla, Khachatryan:2014hpa}. It also includes
a quantification of the difference between the fragmentation of \PW~and Higgs
bosons decaying hadronically.
The fraction of signal events that do not enter any of
the three categories changes from 2\% at 1.1\TeV to 20\% at
3.0\TeV. The uncertainty associated with migration out of the three
categories is estimated to be much smaller than that associated with migration within them.

The uncertainties in the JES ($1-2\%$)~\cite{CMS-DP-2013-011} and
JER (10\%)~\cite{JINST6} impact the signal acceptance in the $\mjj$
counting window.  Each of these systematic contributions provide less than 1\%
uncertainty in the normalization of the expected signal events.

In summary, the uncertainty in the signal normalization associated with
the migration of signal events between categories is larger than the
total contribution of all other uncertainties, which varies from 7\% at
$\mx = 1.1\TeV$ to 15\% at $\mx = 3\TeV$.

The statistical uncertainty in the total background ranges from 15\% at 1.3\TeV up to 100\% at 3\TeV.
It is calculated by
generating pseudo-experiments in the signal and control regions,
assuming Poisson fluctuations in the number of events in each bin
about its central value.
For low $\mjj$, the statistical precision is
limited by the uncertainty in the parameter $N_B$, and for high masses, by
the uncertainty in the slope parameter $a$.
The impact of the choice of the
functional form used in the parameterization of the background
distribution is evaluated by comparing the results from the
exponential fit to those from an alternative power-law function, and is found to
be negligible compared to the statistical uncertainty.

The uncertainty related to the efficiency of the $\nsub$ tagger is
assumed to be fully correlated between the HPLP and LPHP categories
and anticorrelated with the HPHP category. The uncertainties in the
background estimate are uncorrelated between categories, while all
other uncertainties are expected to be fully correlated among all
three categories.

\section{Results}
\label{sec:results}

\begin{figure*}[htbp]
\centering
\includegraphics[width=0.49\textwidth,angle=0]{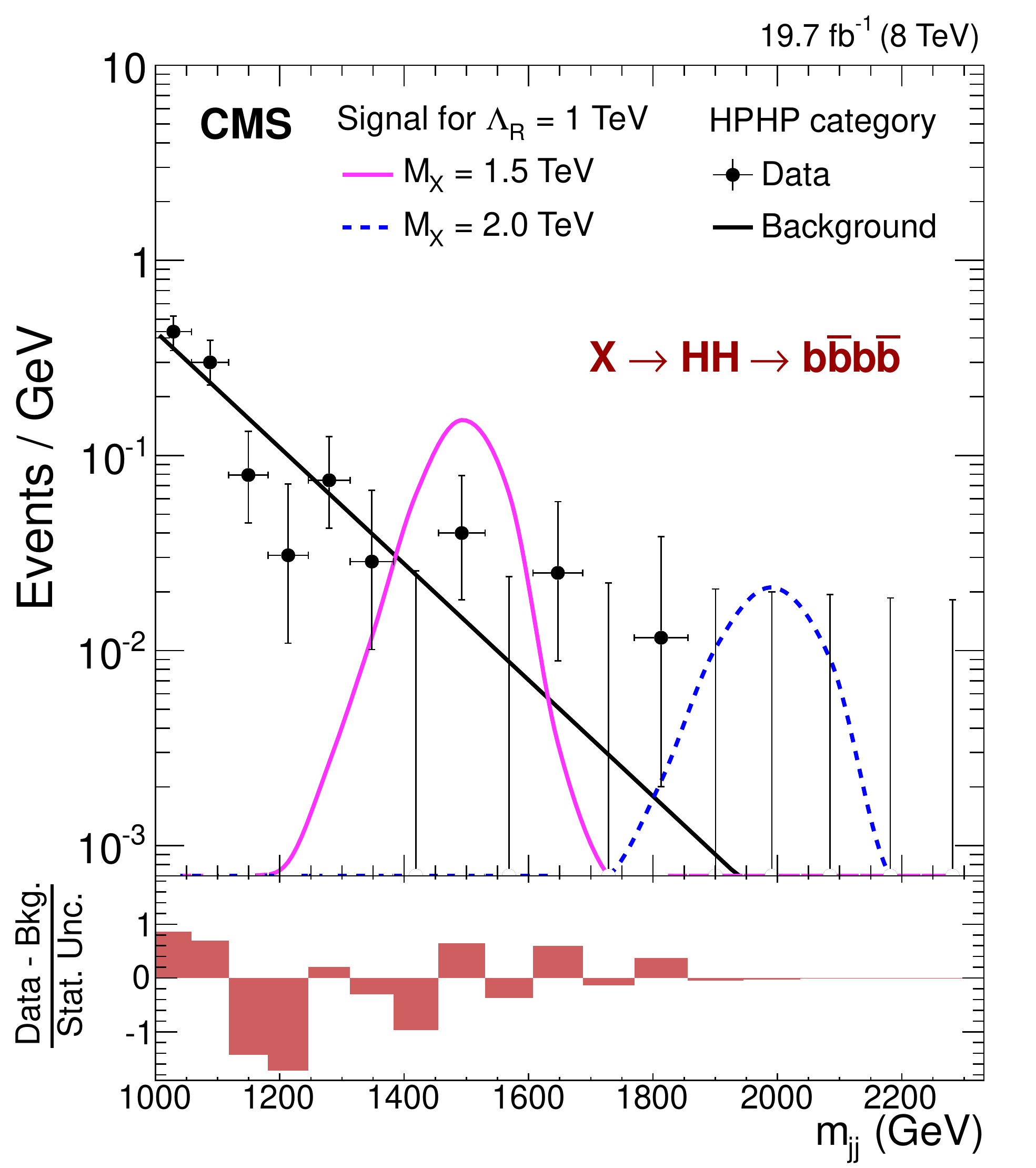}\\
\includegraphics[width=0.49\textwidth,angle=0]{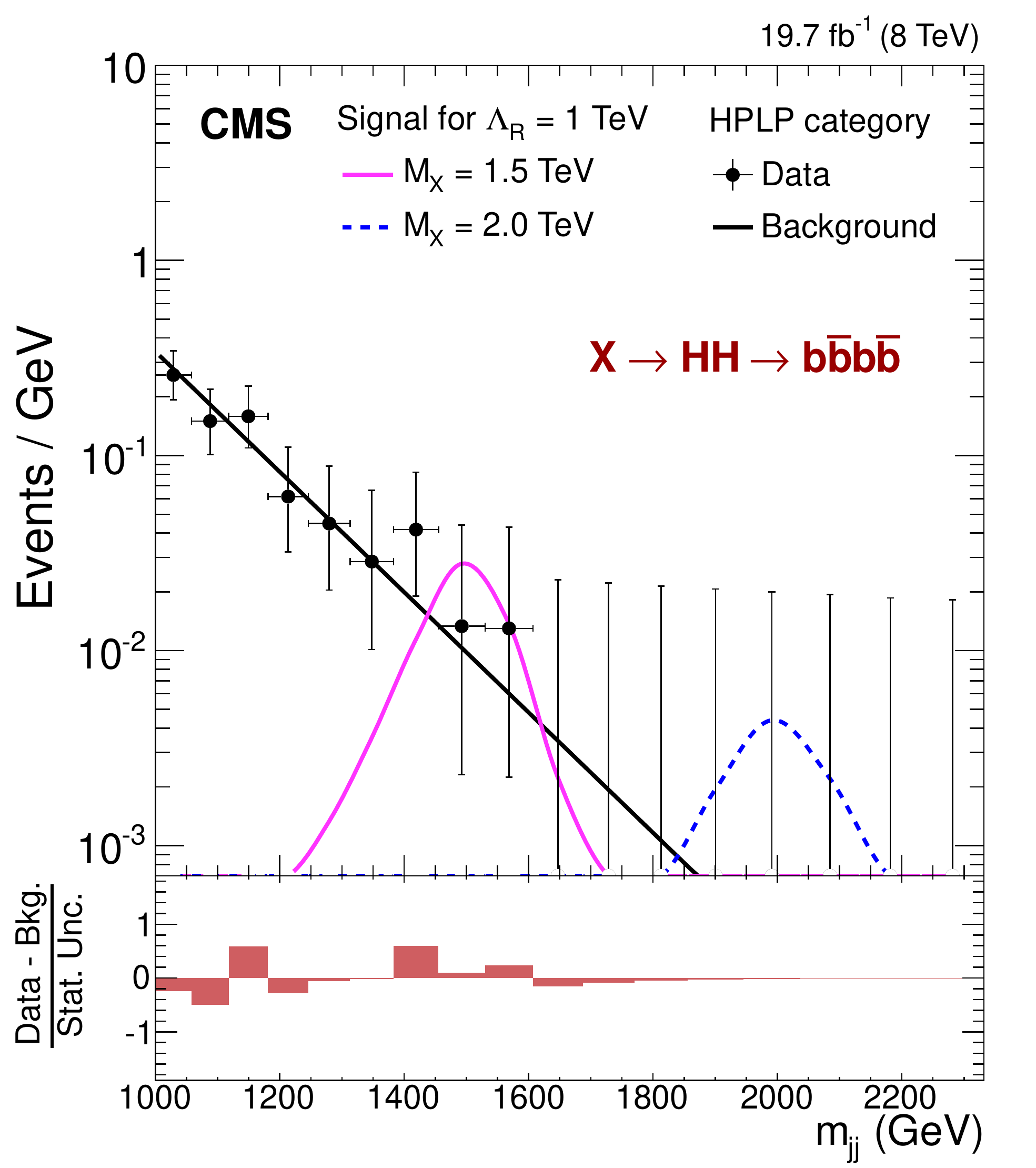}
\includegraphics[width=0.49\textwidth,angle=0]{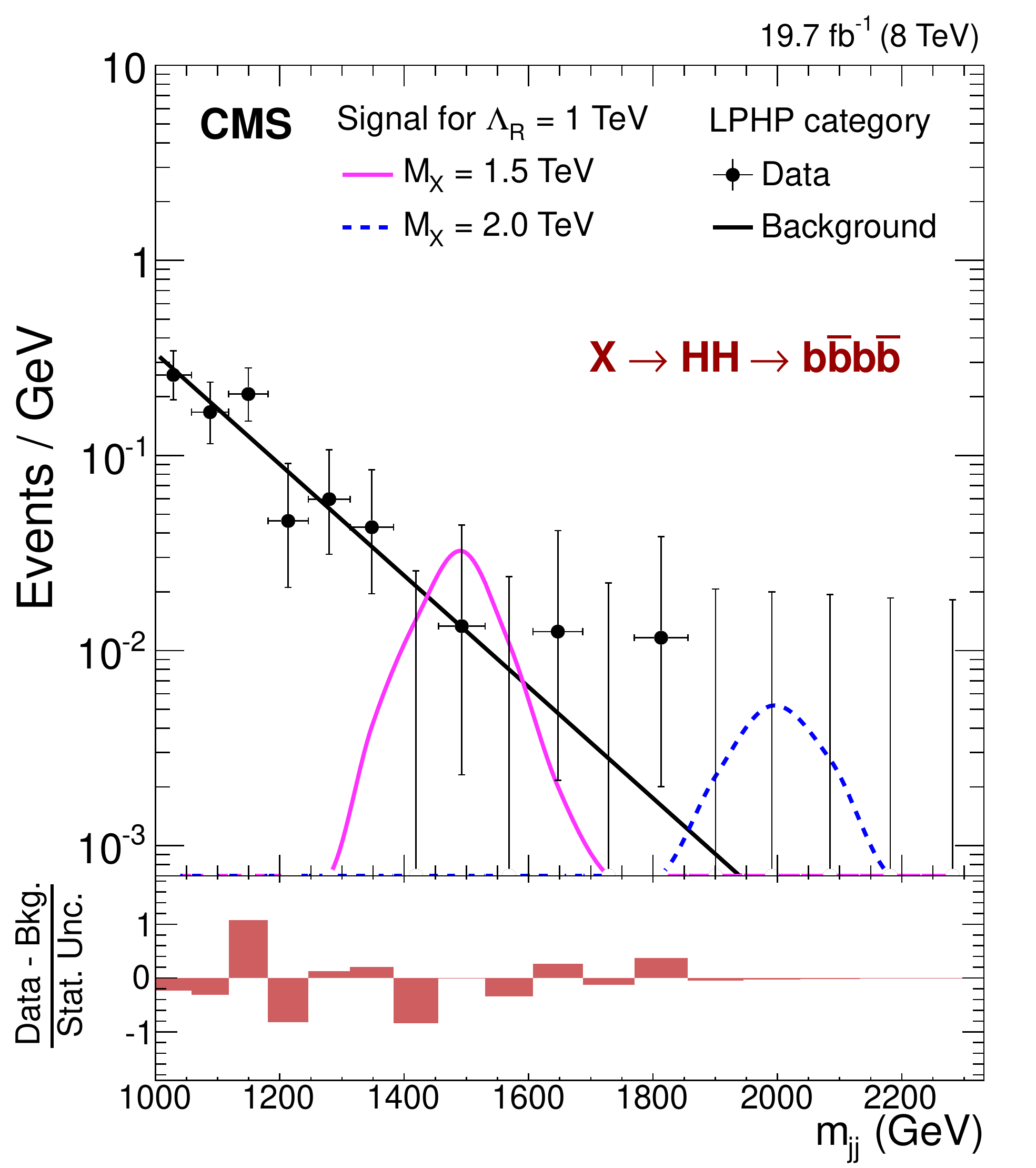}
\caption{Observed $\mjj$ spectrum (black points) compared with a background estimate (black line),
obtained in background only hypothesis, for HPHP (top), HPLP (bottom-left), and LPHP (bottom-right) categories.
The simulated radion resonances of $\mx = 1.5$ and 2\TeV are also shown.
The lower panel in each plot shows the difference between the number of observed and estimated background
events divided by the statistical uncertainty estimated from data.
}
\label{fig:doubleVtagBG}
\end{figure*}

The observed data are shown
separately for the three event categories in Fig.~\ref{fig:doubleVtagBG}. For comparison, we also show the
predictions obtained for the background-only hypothesis. The $N_B$ normalization parameter is extracted for all
events in the signal region with $1 < \mjj < 3\TeV$.  The bottom panel
of each plot shows the difference between the observed data and
the predicted background, divided by the statistical uncertainty
estimated in the data. The background model describes the data within their
statistical uncertainties.
The events with the largest masses in the HPHP,
HPLP, and LPHP categories are at $\mjj = 1780$, 1560, and 1800\GeV,
respectively.

\begin{figure*}[htbp]
\centering
\includegraphics[width=0.49\textwidth]{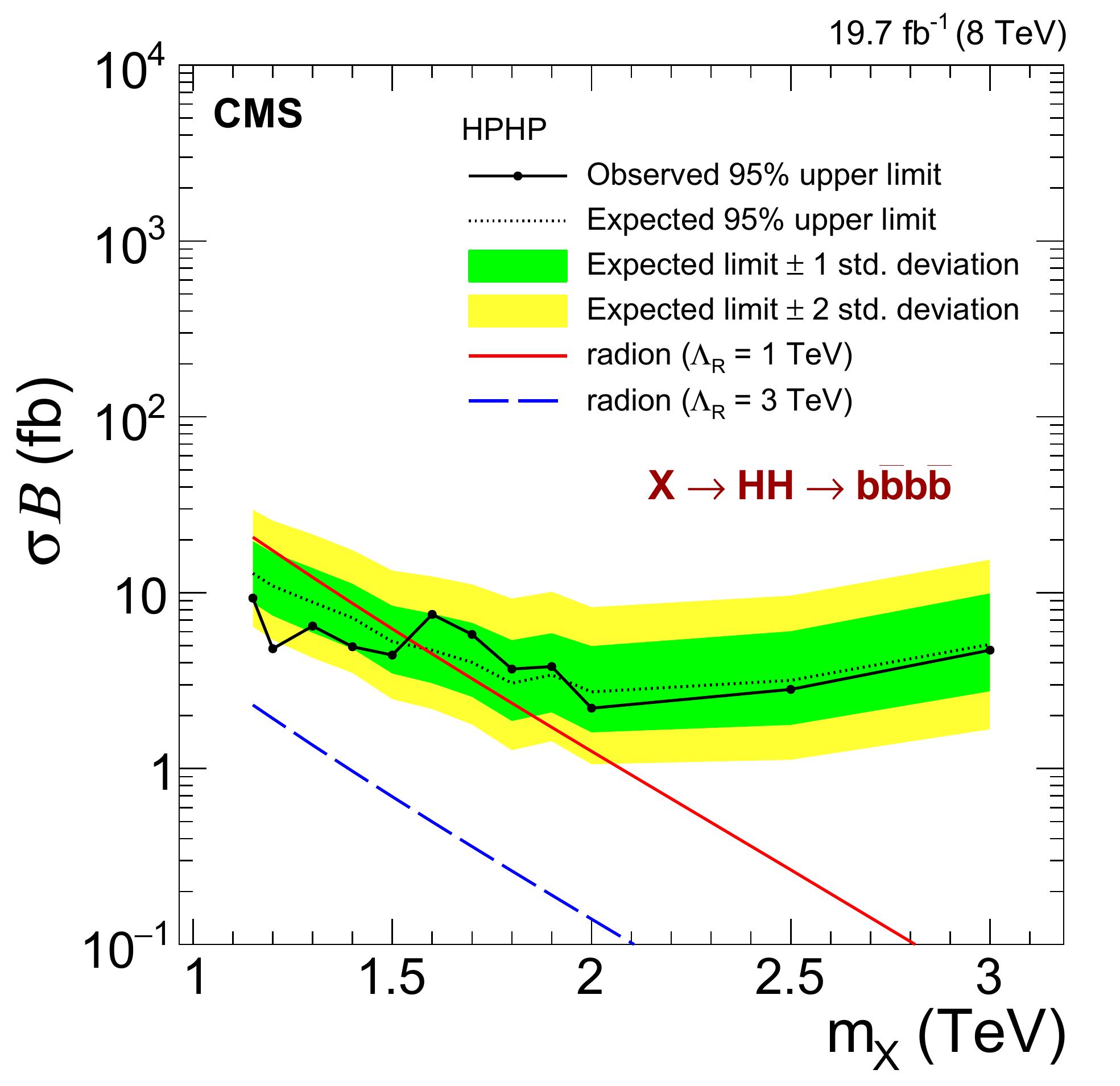}\\
\includegraphics[width=0.49\textwidth]{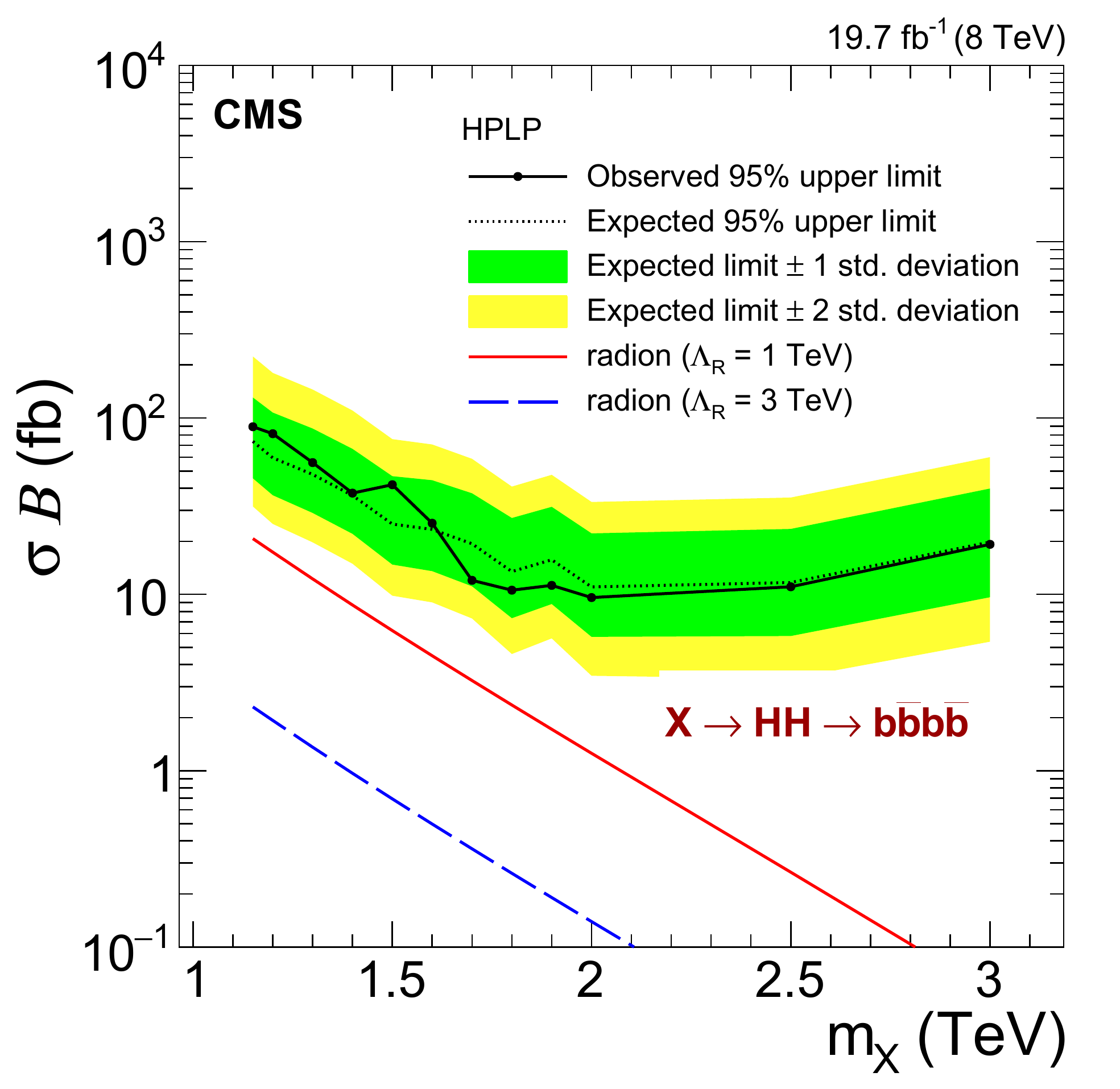}
\includegraphics[width=0.49\textwidth]{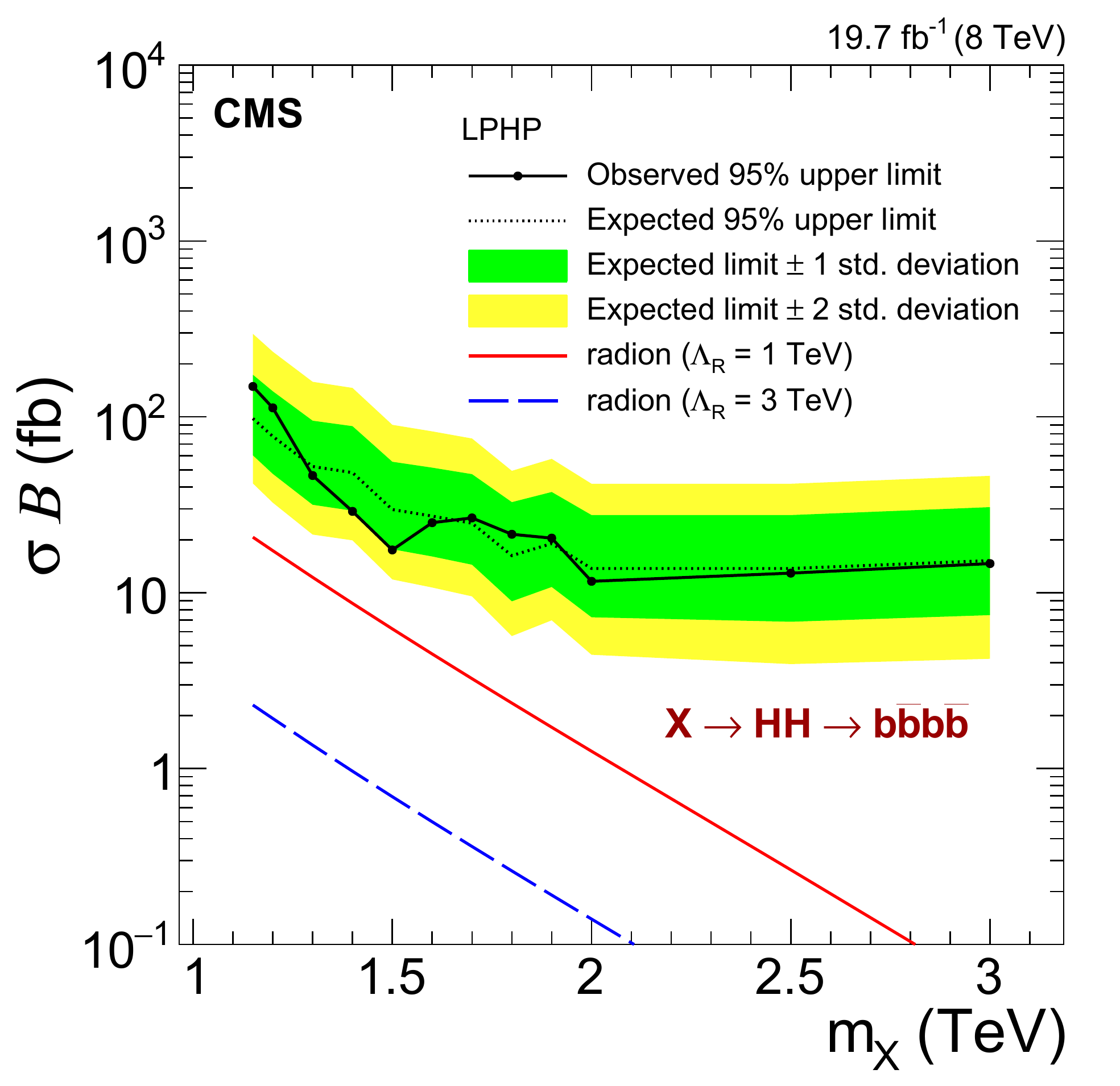}
\caption{Observed and expected 95\% CL upper limits on the product of cross section of a narrow resonance and the
  branching fraction $\sigma(\Pg\Pg \to \PX) \, \mathcal{B}(\PX \to \HH \to \bbbar \bbbar)$.
Theory curves corresponding to WED models with radion are superimposed.}
\label{fig:doubleVtagLimits}
\end{figure*}

\begin{figure}[htb]
\centering
\includegraphics[width=\cmsFigWidth]{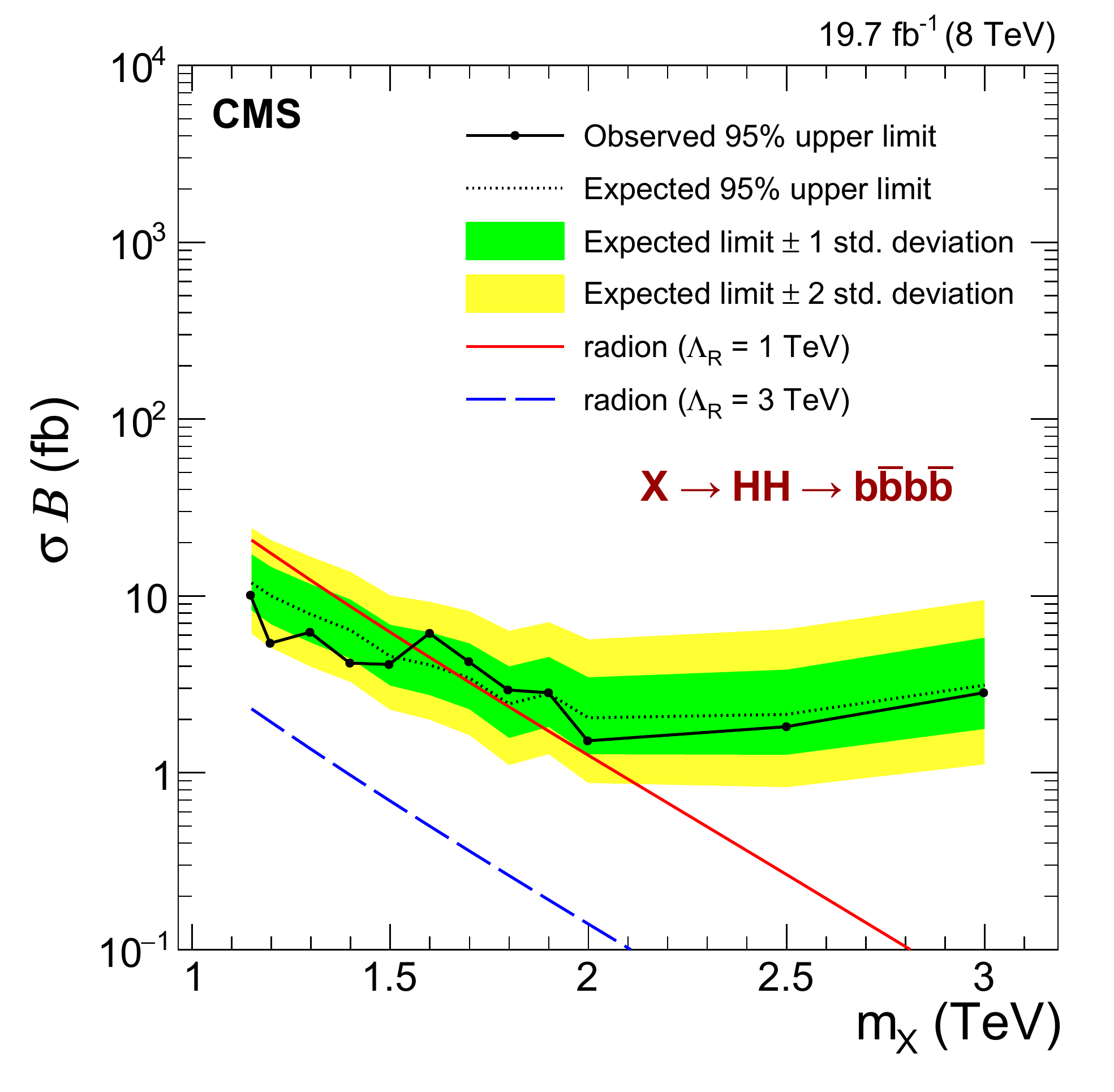}
\caption{
Observed and expected 95\% CL upper limits on the product of cross section of a narrow resonance and the
  branching fraction $\sigma(\Pg\Pg \to \PX) \, \mathcal{B}(\PX \to \HH \to \bbbar \bbbar)$.
Theory curves corresponding to WED models with radion are also shown.
\label{fig:HtagresultsCOMBINED}}
\end{figure}

Upper limits on the cross section for the production of
resonances are extracted using the asymptotic approximation of the
CL$_\mathrm{s}$ method~\cite{CLS1,Junk:1999kv}.
Figure~\ref{fig:doubleVtagLimits} shows the observed and expected 95\%
confidence level (CL) upper limits on the product of the cross section and
the branching fraction $\sigma(\Pg\Pg \to \PX) \, \mathcal{B}(\PX \to \HH \to \bbbar\bbbar)$
obtained for each event category.
The HPHP
category is always the most sensitive, nevertheless above 2\TeV
the HPLP and LPHP categories are also important because of
inefficiencies in N-subjettiness at high \pt.
Figure~\ref{fig:HtagresultsCOMBINED} and
Table~\ref{tab:limitsHighMass} provide the combined limits.
The excluded cross sections at 95\% CL vary from 10\unit{fb} at 1.15\TeV
to 1.5\unit{fb} at 2\TeV. Above 2\TeV the excluded cross sections increase to 2.8\unit{fb} at 3\TeV, since the sensitivity is limited by
the increasing inefficiency of \PH jet identification, as described in
Section \ref{sec:analysis}.

\begin{figure}[htb]
\centering
\includegraphics[width=\cmsFigWidth]{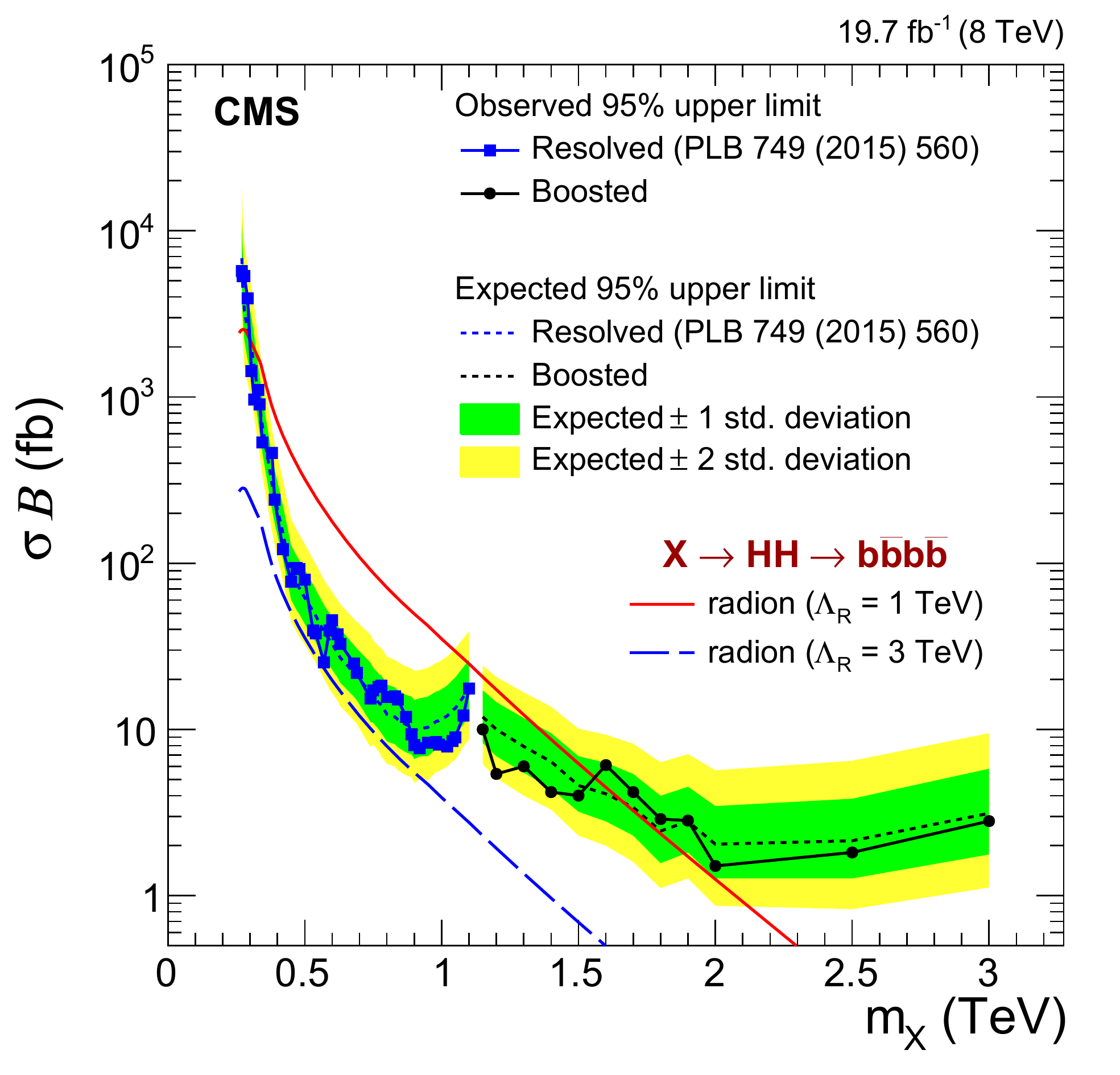}
\caption{
Observed and expected 95\% CL upper limits on the product of cross section of a narrow resonance and the
  branching fraction $\sigma(\Pg\Pg \to \PX) \, \mathcal{B}(\PX \to \HH \to \bbbar \bbbar)$.
Theory predictions corresponding to WED models with a radion are also shown. Results from the resolved analysis of Ref.~\cite{Khachatryan:2015year} are shown by blue squares. For clarity, only a representative subset of the points are provided from the resolved analysis. The result from this paper is shown in black dots.
\label{fig:CommonPlot}}
\end{figure}

Figure \ref{fig:CommonPlot} extends the $\PX \to \HH \to \bbbar\bbbar$ search down to $\mx = 260\GeV$
by including limits from Ref. \cite{Khachatryan:2015year}. This search, referred to as the resolved analysis, considers a case where the decay products from two Higgs bosons are reconstructed
as four jets. It is interesting to observe that the sensitivity of the resolved analysis
starts to degrade at $\mx \approx 1\TeV$. At this point the typical angular distance between two jets from one Higgs boson reaches $\Delta R = 4\mH/\mx \approx 0.5$
and the two jets overlap \cite{Gouzevitch:2013qca}. Above 1.1\TeV the boosted analysis becomes more sensitive.

\begin{table}[htb]
  \topcaption{
    Observed and expected 95\% CL upper limits on the product of cross section and the
    branching fraction $\sigma(\Pg\Pg \to \PX) \, \mathcal{B}(\PX \to \HH \to \bbbar
    \bbbar)$ for HPHP, HPLP and LPHP categories combined.
    The one standard deviation on the 95\% CL upper limit is also provided.\label{tab:limitsHighMass}}
\centering
\begin{tabular}{  c  c  c }
\hline
$\mx$  & Observed limit & Expected limit ${\pm}1\sigma$ \\
(\GeVns{})& (fb) & (fb)\\ \hline\\[-2.2ex]
1150 & 10.0 & 11.9$\; \pm ^{5.3}_{3.6}$\\[0.4ex]
1200 & 5.4 & 10.0$\; \pm ^{4.6}_{3.1}$   \\[0.4ex]
1300 & 6.0 & 7.9$\; \pm ^{3.8}_{2.4}$   \\[0.4ex]
1400 & 4.2 & 6.4$\; \pm ^{3.1}_{2.0}$   \\[0.4ex]
1500 & 4.0 & 4.6$\; \pm ^{2.3}_{1.4}$   \\[0.4ex]
1600 & 6.1 & 4.1$\; \pm ^{2.2}_{1.3}$   \\[0.4ex]
1700 & 4.2 & 3.4$\; \pm ^{2.0}_{1.1}$   \\[0.4ex]
1800 & 2.9 & 2.5$\; \pm ^{1.5}_{0.9}$   \\[0.4ex]
1900 & 2.8 & 2.8$\; \pm ^{1.7}_{1.0}$   \\[0.4ex]
2000 & 1.5 & 2.0$\; \pm ^{1.4}_{0.9}$   \\[0.4ex]
2500 & 1.8 & 2.1$\; \pm ^{1.7}_{0.9}$   \\[0.4ex]
3000 & 2.8 & 3.1$\; \pm ^{2.7}_{1.4}$   \\[0.4ex]
\hline
\end{tabular}
\end{table}

To quantify the sensitivity of this analysis to new physics, the
limits are compared to predictions of radion production for $\LambdaR= 1$ and $3\TeV$, as shown in Fig.~\ref{fig:HtagresultsCOMBINED}.  We
find that a radion corresponding to $\LambdaR = 1\TeV$ is excluded by the boosted analysis alone, for
masses between 1.15 and 1.55\TeV. This result extends the limits already set by the resolved analysis from 0.3 to 1.1 \TeV.

\section{Summary}
\label{sec:conclusions}

A search is presented for narrow heavy resonances decaying into a pair
of Higgs bosons in proton-proton collisions collected by the CMS
experiment at $\sqrt{s}=8\TeV$. The full data sample of \intlumi is
explored.  The background from multijet and \ttbar events is
significantly reduced by applying requirements related to the flavor
of the jet, its mass, and its substructure.
No significant excess of events is observed above the background expected from the SM processes.
The results are
interpreted as exclusion limits at 95\% confidence on the
production cross section for $\mx$ between 1.15 and 3.0\TeV,
extending significantly beyond 1.5 TeV the reach of previous searches.
A radion with scale parameter $\LambdaR = 1\TeV$ decaying into $\PH\PH$ is excluded for $1.15 < \mx <1.55\TeV$ for the first time in direct searches.

\begin{acknowledgments}
We congratulate our colleagues in the CERN accelerator departments for the excellent performance of the LHC and thank the technical and administrative staffs at CERN and at other CMS institutes for their contributions to the success of the CMS effort. In addition, we gratefully acknowledge the computing centers and personnel of the Worldwide LHC Computing Grid for delivering so effectively the computing infrastructure essential to our analyses. Finally, we acknowledge the enduring support for the construction and operation of the LHC and the CMS detector provided by the following funding agencies: BMWFW and FWF (Austria); FNRS and FWO (Belgium); CNPq, CAPES, FAPERJ, and FAPESP (Brazil); MES (Bulgaria); CERN; CAS, MoST, and NSFC (China); COLCIENCIAS (Colombia); MSES and CSF (Croatia); RPF (Cyprus); MoER, ERC IUT and ERDF (Estonia); Academy of Finland, MEC, and HIP (Finland); CEA and CNRS/IN2P3 (France); BMBF, DFG, and HGF (Germany); GSRT (Greece); OTKA and NIH (Hungary); DAE and DST (India); IPM (Iran); SFI (Ireland); INFN (Italy); MSIP and NRF (Republic of Korea); LAS (Lithuania); MOE and UM (Malaysia); CINVESTAV, CONACYT, SEP, and UASLP-FAI (Mexico); MBIE (New Zealand); PAEC (Pakistan); MSHE and NSC (Poland); FCT (Portugal); JINR (Dubna); MON, RosAtom, RAS and RFBR (Russia); MESTD (Serbia); SEIDI and CPAN (Spain); Swiss Funding Agencies (Switzerland); MST (Taipei); ThEPCenter, IPST, STAR and NSTDA (Thailand); TUBITAK and TAEK (Turkey); NASU and SFFR (Ukraine); STFC (United Kingdom); DOE and NSF (USA).

Individuals have received support from the Marie-Curie program and the European Research Council and EPLANET (European Union); the Leventis Foundation; the A. P. Sloan Foundation; the Alexander von Humboldt Foundation; the Belgian Federal Science Policy Office; the Fonds pour la Formation \`a la Recherche dans l'Industrie et dans l'Agriculture (FRIA-Belgium); the Agentschap voor Innovatie door Wetenschap en Technologie (IWT-Belgium); the Ministry of Education, Youth and Sports (MEYS) of the Czech Republic; the Council of Science and Industrial Research, India; the HOMING PLUS program of the Foundation for Polish Science, cofinanced from European Union, Regional Development Fund; the OPUS program of the National Science Center (Poland); the Compagnia di San Paolo (Torino); MIUR project 20108T4XTM (Italy); the Thalis and Aristeia programs cofinanced by EU-ESF and the Greek NSRF; the National Priorities Research Program by Qatar National Research Fund; the Rachadapisek Sompot Fund for Postdoctoral Fellowship, Chulalongkorn University (Thailand); the Chulalongkorn Academic into Its 2nd Century Project Advancement Project (Thailand); and the Welch Foundation, contract C-1845.
\end{acknowledgments}
\bibliography{auto_generated}

\cleardoublepage \appendix\section{The CMS Collaboration \label{app:collab}}\begin{sloppypar}\hyphenpenalty=5000\widowpenalty=500\clubpenalty=5000\textbf{Yerevan Physics Institute,  Yerevan,  Armenia}\\*[0pt]
V.~Khachatryan, A.M.~Sirunyan, A.~Tumasyan
\vskip\cmsinstskip
\textbf{Institut f\"{u}r Hochenergiephysik der OeAW,  Wien,  Austria}\\*[0pt]
W.~Adam, E.~Asilar, T.~Bergauer, J.~Brandstetter, E.~Brondolin, M.~Dragicevic, J.~Er\"{o}, M.~Flechl, M.~Friedl, R.~Fr\"{u}hwirth\cmsAuthorMark{1}, V.M.~Ghete, C.~Hartl, N.~H\"{o}rmann, J.~Hrubec, M.~Jeitler\cmsAuthorMark{1}, A.~K\"{o}nig, M.~Krammer\cmsAuthorMark{1}, I.~Kr\"{a}tschmer, D.~Liko, T.~Matsushita, I.~Mikulec, D.~Rabady, N.~Rad, B.~Rahbaran, H.~Rohringer, J.~Schieck\cmsAuthorMark{1}, R.~Sch\"{o}fbeck, J.~Strauss, W.~Treberer-Treberspurg, W.~Waltenberger, C.-E.~Wulz\cmsAuthorMark{1}
% ...
\vskip\cmsinstskip
% ...
1:~~Also at Vienna University of Technology, Vienna, Austria\\
% ...

\end{sloppypar}
\end{document}